\documentclass[prl,aps,twocolumn,superscriptaddress,showpacs,
preprintnumbers,amsmath,amssymb]{revtex4}
\usepackage{graphicx}
\begin{document}
%\preprint{PRL}
\title{Orbital excitations in titanates}
\author{C.~Ulrich}
\affiliation{Max-Planck-Institut~f\"{u}r~Festk\"{o}rperforschung,
Heisenbergstr.~1, D-70569 Stuttgart, Germany}
\author{A.~G\"{o}ssling}
\affiliation{II. Physikalisches Institut, Universit\"{a}t zu K\"{o}ln,
50937 K\"{o}ln, Germany}
\author{M.~Gr\"{u}ninger}
\affiliation{II. Physikalisches Institut, Universit\"{a}t zu K\"{o}ln,
50937 K\"{o}ln, Germany}
\author{M.~Guennou}
\affiliation{Max-Planck-Institut~f\"{u}r~Festk\"{o}rperforschung,
Heisenbergstr.~1, D-70569 Stuttgart, Germany}
\author{H.~Roth}
\affiliation{II. Physikalisches Institut, Universit\"{a}t zu K\"{o}ln,
50937 K\"{o}ln, Germany}
\author{M.~Cwik}
\affiliation{II. Physikalisches Institut, Universit\"{a}t zu K\"{o}ln,
50937 K\"{o}ln, Germany}
\author{T.~Lorenz}
\affiliation{II. Physikalisches Institut, Universit\"{a}t zu K\"{o}ln,
50937 K\"{o}ln, Germany}
\author{G.~Khaliullin}
\affiliation{Max-Planck-Institut~f\"{u}r~Festk\"{o}rperforschung,
Heisenbergstr.~1, D-70569 Stuttgart, Germany}
\author{B.~Keimer}
\affiliation{Max-Planck-Institut~f\"{u}r~Festk\"{o}rperforschung,
Heisenbergstr.~1, D-70569 Stuttgart, Germany}

\date{\today}

\begin{abstract}
Raman scattering is used to observe pronounced electronic excitations 
around 230 meV -- well above the two-phonon range -- in the Mott insulators 
LaTiO$_3$ and YTiO$_3$. Based on the temperature, polarization, and photon 
energy dependence, the modes are identified as orbital excitations. The 
observed profiles bear a striking resemblance to magnetic Raman modes in
the insulating parent compounds of the superconducting cuprates, indicating 
an unanticipated universality of the electronic excitations in transition 
metal oxides.
\end{abstract}


\maketitle

Transition metal oxides with orbital degeneracy exhibit a host of
intriguing physical properties, such as ``colossal magnetoresistance" 
in manganites or unconventional superconductivity in ruthenium oxides 
\cite{Ima98,Tok00}. Collective oscillations of the valence electrons 
between different atomic orbitals (termed ``orbitons") contain a
wealth of information about the microscopic interactions underlying 
this behavior. Experiments introducing Raman scattering as a direct 
probe of orbitons in LaMnO$_3$ have hence opened up new perspectives 
for a quantitative understanding of colossal magnetoresistance \cite{Sai01}. 
The results, however, have proven to be quite controversial, because the 
modes observed in LaMnO$_3$ are difficult to discriminate from ordinary
two-phonon excitations \cite{Gru02,Sai02}.

Insulating titanates such as LaTiO$_3$ and YTiO$_3$ have only one valence 
electron residing on the Ti$^{3+}$ ions. Despite their similar, nearly 
cubic lattice structures, the magnetic ground states of LaTiO$_3$ and 
YTiO$_3$ are quite different: Whereas LaTiO$_3$ orders antiferromagnetically 
at $\rm T_N \sim 150$ K, YTiO$_3$ is a ferromagnet with Curie temperature 
$\rm T_C \sim 30$ K. The origin of this difference lies in the large degeneracy 
of the quantum states available to the valence electron. In addition to
its spin degeneracy, this electron can occupy any combination of the three 
$t_{2g}$-orbitals $xy$, $xz$ and $yz$. The six single-ion states on neighboring
sites are coupled by the superexchange interaction and collective lattice 
distortions, which generate a plethora of nearly degenerate many-body states 
with different spin and orbital ordering patterns.

The mechanisms selecting the ground state out of this large manifold of states 
have recently been a focus of intense research. Two theoretical approaches 
have emerged. According to point charge model \cite{Cwi03,Moc03,Sch04}, 
band structure \cite{Saw97,Sol04} and LDA-DMFT \cite{Pav04} calculations, 
the degeneracy of the single-ion $t_{2g}$-levels is lifted by subtle lattice 
distortions. These distortions favor linear combinations of type $(xz+yz+xy)/\sqrt3$
for LaTiO$_3$ and a staggered pattern of planar orbitals of the kind $(xz+xy)/\sqrt2$ 
in YTiO$_3$. This proposal has received support from NMR \cite{Ito99,Kiy03} and 
neutron diffraction \cite{Aki01} experiments. However, the small ordered moment
and some aspects of the low-temperature spin excitation spectra, such as their 
spatial isotropy and small gaps, are difficult to reconcile with the scenario 
of static, lattice-driven orbital order \cite{Kei00,Ulr02}. An alternative 
theoretical approach \cite{Kha00,Kha02} inspired by these findings emphasizes 
the collective quantum dynamics of the orbitals, driven by intersite correlations 
via the superexchange interaction. Frustrations inherent to these interactions give 
rise to strong quantum fluctuations that suppress orbital ordering and stabilize 
more isotropic charge distributions around the Ti$^{3+}$ ions.

Orbital excitations have the potential to discriminate between
these conflicting scenarios. When the orbital dynamics is quenched
by lattice distortions, one expects a number of transitions
between well-defined crystal field levels, with selection rules
dictated by the symmetry of these distortions. The picture is
similar to the local crystal field excitations of, e.g., Ti$^{3+}$
impurities in a distorted corundum lattice \cite{Nel67}. In an
orbitally fluctuating state, on the other hand, the orbital
excitations are collective modes with selection rules controlled
by many-body effects and hence are very different from local crystal
field excitations. Therefore, the observation of
orbital excitations and comparison of their properties in YTiO$_3$
and LaTiO$_3$, two materials with different lattice distortions
and magnetic ground states, is of crucial importance for the
understanding of this hotly debated issue.

We have used Raman scattering to search for orbital excitations in 
LaTiO$_3$ and YTiO$_3$. The samples were high quality single crystals 
with $\rm T_N$ = 146 K and $\rm T_C$ = 27 K, respectively, grown by 
the floating zone technique described in Ref.~\cite{Cwi03}. The LaTiO$_3$ 
crystal was partly twinned. Raman measurements were performed using 
a Dilor-XY triple spectrometer equipped with a CCD-detector. The 
measured spectral intensity was calibrated by using a white light source 
(Ulbricht--sphere) as reference. Different lines of an Ar$^+$/Kr$^+$ 
mixed--gas ion laser were used. In order to avoid heating of the sample, 
the power of the incident laser beam was kept below 10 mW at the sample
position. The experiments were performed in backscattering geometry 
parallel to the crystallographic $b$--direction (within the P$bnm$ space group).
The polarizations of the incident and scattered photons are specified
as ($\alpha,\beta$) where $\alpha$ and $\beta$ denote the $z,x,z'$ or
$x'$ directions. Note that $z\parallel c$ is along the nearest-neighbor 
Ti--Ti bond whereas $x$ is along the next-nearest-neighbor Ti--Ti
direction in the $ab$-plane. The $z'$ and $x'$ directions are rotated 
by 45$^\circ$ from $z$ and $x$.

%%%%%%%%%%% FIG.1: %%%%%%%%%%%%%%%%%%%
\begin{figure}
\includegraphics[width=0.8\linewidth]{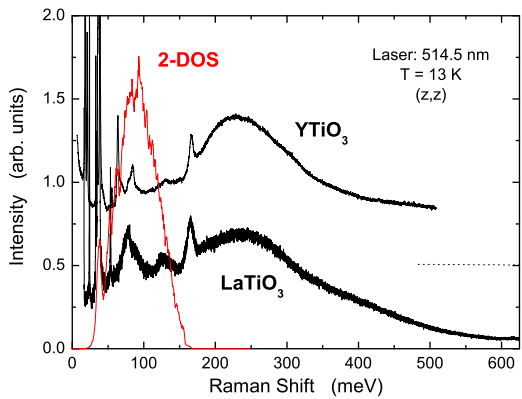}
\vspace*{-0.2cm}
\caption{\label{fig_orb}
Raman spectra of LaTiO$_3$ and YTiO$_3$ measured at T = 13 K using the 
514.5 nm laser line in $(z,z)$ polarization. The red line shows the calculated 
two--phonon density of states.}
\end{figure}
%%%%%%%%%%% FIG.1: %%%%%%%%%%%%%%%%%%%
Figure 1 shows the Raman spectra of LaTiO$_3$ and YTiO$_3$ over a wide energy 
range.  Up to 80~meV, the spectra are dominated by one-phonon excitations reported 
before \cite{Ree97,Ili04}. A series of weak features extends up to 170 meV. At 
still higher frequencies, around 235 meV, a pronounced broad peak is observed
in both compounds. The latter feature has hitherto not been reported and 
constitutes the central observation of this paper.

In order to establish whether the broad peak can be attributed to
two-phonon excitations, we have evaluated the two-phonon density of 
states (2-DOS) of YTiO$_3$ by convoluting the one-phonon dispersion 
curves calculated within a shell model whose parameters were fitted to
the observed phonon frequencies \cite{Ulr04}. The 2-DOS,
responsible for the weak features between 80 and 170 meV,
terminates at about 170 meV. The broad peak thus occurs well above 
the two-phonon spectral range, in contrast to the features around 
160 meV discussed for LaMnO$_3$ \cite{Sai01,Gru02}. Since higher
phonon orders are expected to appear even weaker, lattice vibrations 
can be ruled out as the origin of the high-energy Raman peak. 
As an aside, we note the presence of a sharp feature at the upper 
edge of the 2-DOS around 165 meV, which is not reproduced by our 
calculation. This peak is also observed by inelastic neutron scattering
\cite{Ulr04}. A similar Raman mode is also observed in other
transition metal oxides \cite{Sai01,Lyo88,Yos92,Ree97}, and has
been attributed to a two-phonon excitation
\cite{Gru02,Lyo88,Yos92,Ree97}.

%%%%%%%%%%%% FIG.2: resonance %%%%%%%%%%%%%%%%%%%%%%%%%%%%%%%
\begin{figure}
\includegraphics[width=0.8\linewidth]{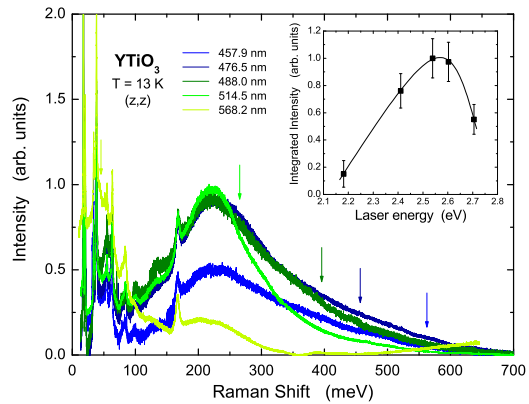}
\vspace*{-0.3cm}
\caption{\label{fig_orb_res}
Raman spectrum of YTiO$_3$ measured at T = 13 K for different laser
lines. A frequency independent background has been subtracted from
every profile. The arrows depict the position of the
photoluminescence peak at 2.14 eV for the different laser lines.
The inset shows the integrated intensity of the broad high-energy
peak.}
\end{figure}
%%%%%%%%%%%% FIG.2: resonance %%%%%%%%%%%%%%%%%%%%%%%%%%%%%%%
In order to rule out photoluminescence as a cause of the broad
peak, we have repeated the Raman measurements with different laser
lines between 457.9 nm and 568.2 nm (Fig. 2). The peak at 235 meV
does not shift as the laser frequency is changed, ruling out
photoluminescence as origin and demonstrating that the peak position
corresponds to a genuine excitation frequency. The data also show 
the manifestations of a weak photoluminescence feature of energy 
2.14 eV and width 75 meV, which is responsible for the low-energy 
intensity in the data at the laser wavelength 568.2 nm.
Since the Raman spectra are measured at energy transfers relative
to the incident laser energy, the luminescence peak is shifted
when the laser energy is changed and contributes to the variation
of the apparent lineshape of the 235 meV mode with laser
wavelength. However, the main part of the mode profile is
unaffected by this extraneous feature. The inset of Figure 2 shows
that the integrated spectral weight of the 235 meV mode in
YTiO$_3$ exhibits a pronounced resonant enhancement at a laser
frequency of $\sim 2.54$ eV. The Raman spectra displayed in Figure
3 demonstrate that its spectral weight is also strongly
temperature dependent, decreasing continuously with increasing
temperature.

%%%%%%%%%%%% FIG.3: temperature %%%%%%%%%%%%%%%%%%%%
\begin{figure}
\includegraphics[width=0.8\linewidth]{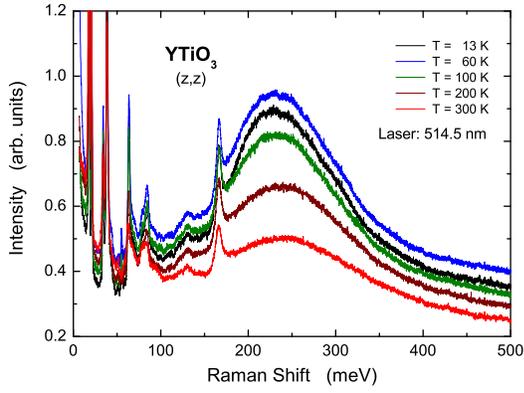}
\vspace*{-0.2cm}
\caption{\label{fig_orb_Tdep}
Temperature dependence of the Raman spectrum of YTiO$_3$ in $(z,z)$ 
geometry, using the 514.5 nm laser line.}
\end{figure}
%%%%%%%%%%%% FIG.3: temperature %%%%%%%%%%%%%%%%%%%%
The intensity of the mode depends strongly on the light polarization 
(Fig. 4). In a cubic crystal, the Raman intensity for parallel and 
crossed polarizations can be expressed via $E_g, T_{2g}$ and $A_{1g}$ 
symmetry components:
\begin{eqnarray}
I_{parallel}&=&(\frac{2}{3}-a_{\theta}-b_{\theta})E_g+
(a_{\theta}+b_{\theta})T_{2g}+\frac{1}{3}A_{1g}, \nonumber \\
I_{crossed}&=&b_{\theta}E_g+(\frac{1}{2}-b_{\theta})T_{2g},
\end{eqnarray}
where $\theta$ is the angle between the incident electric field
vector and the $c$-axis, $a_{\theta}=\frac{1}{2}\sin^2 \theta$,
$b_{\theta}=\frac{3}{8}\sin^2 2\theta$, and $\theta=90^{\circ}$
corresponds to the $a$ (or $b$) axis. These relations provide a 
good description of the observed polarization dependences in both 
compounds (insets in Fig. 4). Possible corrections induced by 
lattice distortions away from cubic symmetry therefore
appear to be rather small. The observed nearly cubic symmetry of
the high-energy Raman response is in accord with the cubic
symmetry of spin wave dispersions in these compounds
\cite{Kei00,Ulr02}.

We now turn to the interpretation of the 235 meV mode. Having
ruled out phonons and photoluminescence, we attribute the mode 
to electronic scattering. Its lineshape, temperature, and 
polarization dependence shows striking similarities (elaborated 
later on) with the Raman scattering from spin-pair excitations 
in the insulating parent compounds of the superconducting cuprates 
\cite{Lyo88,Kno90,Yos92}. However, a simple two-magnon origin of 
the broad band is ruled out here because YTiO$_3$ is ferromagnetic, 
and because the peak energy is much higher than twice the magnon 
energies found by neutron scattering: $\sim$ 20 meV in YTiO$_3$ 
\cite{Ulr02} and $\sim$ 44 meV in LaTiO$_3$ \cite{Kei00}. We thus 
attribute the broad Raman mode to the orbital degrees of freedom of
the valence electron.

%%%%%%%%%%%%% FIG.4: polarization %%%%%%%%%%%%%%%%%%%%%%%%%
\begin{figure}
\includegraphics[width=0.72\linewidth]{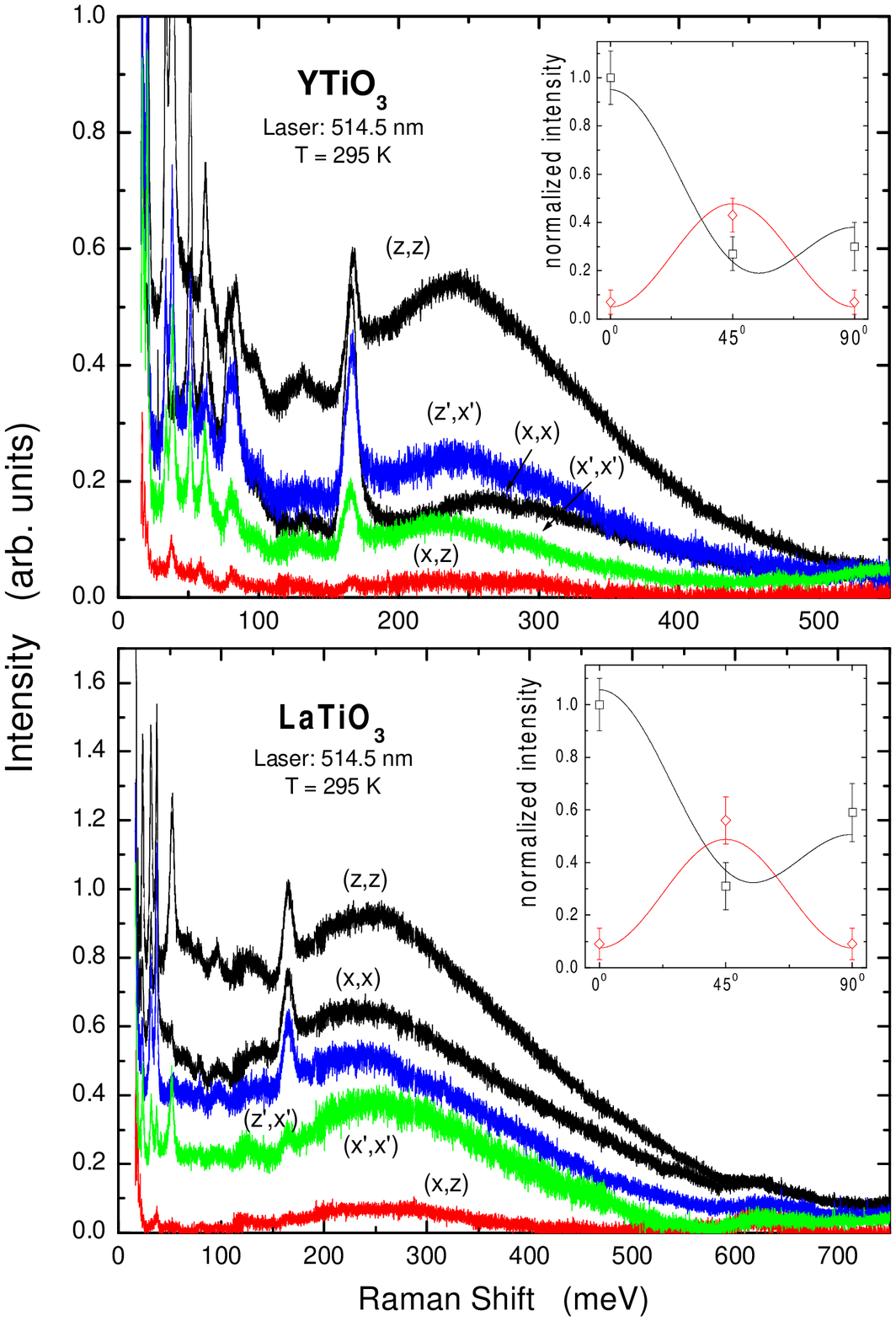}
\vspace*{-0.2cm} \caption{\label{fig_orb_pol} Polarization
dependences of the Raman spectra of YTiO$_3$ and LaTiO$_3$ at room
temperature. The intensity is symmetric with respect to
$(\alpha,\beta) \rightarrow (\beta,\alpha)$. A frequency
independent background has been subtracted.
Insets: scattering intensities for the parallel ($\Box$) and
crossed ($\Diamond$) polarizations as function of angle $\theta$.
Solid and dashed lines are obtained from Eq.~(1), with relative
intensities $E_g:A_{1g}:T_{2g} \approx 1:0.3:0.1$ $(1:0.5:0.1)$ in
YTiO$_3$ (LaTiO$_3$).}
\end{figure}
%%%%%%%%%%%%% FIG.4: polarization %%%%%%%%%%%%%%%%%%%%%%%%%
Let us start with a simple crystal field picture where the
incident photon excites an oxygen $2p$ electron to unoccupied
$t_{2g}$-orbitals via the charge-transfer gap $\Delta_{pd}$ and,
in a second step, the $d$-electron in the ground state orbital
relaxes to the oxygen site emitting a photon and leaving a
``flipped" orbital behind. Predictions for the energy cost of such
a local orbital flip vary between 27 and 240 meV \cite{Cwi03,Pav04,Sol04}, 
but some model calculations yield energies comparable to those observed 
experimentally \cite{Cwi03}. The large width of the crystal field 
transitions can be understood as a consequence of coupling to lattice
vibrations. Other aspects of the Raman data are, however,
difficult to reconcile with a local crystal field scenario. First,
the experimentally observed strong temperature dependence of the
spectral weight is hard to understand within this picture, as the
local $pd$ charge transfer process is not sensitive to temperature
variations. Second, the above described scenario cannot explain
the observed strong resonance at a photon energy of 2.54 eV, which
is far below the $pd$-excitation energy $\Delta_{pd}>4$~eV \cite{Oki95}.

The resonance energy yields an important clue to the microscopic
origin of the broad Raman peak. As it falls in the range of intersite 
$d_i\rightarrow d_j$ excitations in optical spectra \cite{Oki95}, {\it two}
Ti sites must be involved in the Raman process. This means that the 
$t_{2g}$-electron is first transferred to a neighboring site by the 
incident photon that matches the $d_i\rightarrow d_j$ transition energy. 
The intermediate state then relaxes back, leaving behind flipped orbitals 
either on one or on both sites, depending on the actual orbital pattern. 
Consider again the lattice-driven orbital picture, where detailed 
predictions are available \cite{Ish04}. According to these predictions, 
the orbital excitations probed by light polarized along different crystal 
axes are fundamentally different. In particular, the selection rules for 
the four-sublattice pattern $(xz\pm xy)/\sqrt2~,(yz\pm xy)/\sqrt2$ suggested 
for YTiO$_3$ \cite{Saw97,Ito99,Aki01} allow a single orbital-flip for 
$ab$-plane polarization, but only two-orbiton excitations may occur in 
case of light polarization along the $c$-axis \cite{Ish04}. This selection 
rule (which results from a mirror symmetry of the Ti-O-Ti bonds along the
$c$-axis) is in sharp contrast with the observed nearly cubic
selection rules. Similar considerations can be made for the
orbital state proposed for LaTiO$_3$.

The insensitivity of the orbiton excitations to details of the
lattice distortions suggests that electronic interactions play a
major part in driving the orbital state, as proposed in Refs.
\cite{Kei00,Ulr02,Kha00,Kha02}. In this picture, the orbitals
are regarded as quantum objects much like the spins of the correlated 
electrons, and interact with each other via the superexchange mechanism. 
The superexchange originates from virtual charge fluctuations
representing the residual kinetic energy of electrons in a Mott
insulator. Similar to the spin exchange (permutation) operator
$\hat{P}_s=(2 \vec{s}_i\vec{s}_j+\frac{1}{2}n_in_j)$, the orbital
exchange is described by $\hat{P}_{orb}=(2
\vec{\tau}_i\vec{\tau}_j+\frac{1}{2}n_in_j)^{(\gamma)}$, where the
action of pseudospins $\vec{\tau}^{(\gamma)}$ on the $t_{2g}$
orbital triplet depends on the bond direction $\gamma$,
leading to strong frustration of orbitals on a cubic lattice
\cite{Kha00,Kha02}. These frustrations, as well as the
simultaneous spin/orbital exchange described by a product
$\hat{P}_s\hat{P}_{orb}$, result in a large degeneracy of
competing many-body states, including the near-degeneracy of spin
ferro- and antiferromagnetic states. The high-energy response of
orbitals in titanates is governed by the universal scale
$J_{SE}\propto 4t^2/U_{dd}$ ({\it per bond}), controlled by the kinetic
energy $t$ and Hubbard correlation $U_{dd}$. Weak lattice
distortions tend to polarize the many-body orbital state, but the
high-energy orbital fluctuations respect the nearly cubic symmetry
of the electronic ground state. This mechanism has also been
invoked to explain the isotropic spin-wave spectra
\cite{Kei00,Ulr02}.

In the superexchange-driven orbital state of YTiO$_3$
\cite{Kha02}, the light scattering from fluctuations of the
{\it orbital-exchange} bonds, described by $\hat{P}_{orb}$, leads
to a two-orbiton Raman band of cubic $E_g$ symmetry \cite{Kha05},
as observed. The $A_{1g}$ and $T_{2g}$ components are explained
due to next-nearest-neighbor orbital exchange terms that are
enhanced by resonant scattering, in complete analogy to the theory
of magnetic Raman scattering in the cuprates \cite{Sha90}. From an
analysis of the spin-wave data of Ref. \cite{Ulr02}, the highest
orbiton energy was determined as $\sim 2 r_1J_{SE}$ with
$r_1J_{SE}\sim 60$~meV and $r_1 \sim 1.5$ \cite{Kha02}, in
reasonable agreement with a two-orbiton peak at the experimental
value of 235 meV.

Since the orbital excitations are strongly damped because of
frustrating interactions, their response is broad in both momentum
and energy, as observed. The strong enhancement of the
peak intensity at low temperature can be understood if the intermediate
electronic state involved in the resonant Raman process has
high-spin configuration. The relevant matrix element is thus
controlled by intersite spin correlations via the operator
$(\vec{s}_i\vec{s}_j+\frac{3}{4})$, selecting a triplet state. The
square of this quantity changes by roughly factor of two between
the high and low temperature limits, consistent with the experiment.

In summary, the high-energy orbital excitations in LaTiO$_3$ and
YTiO$_3$ are quite insensitive to distortions of the lattice away
from cubic symmetry. The superexchange-driven quantum orbital picture
provides a consistent description of this observation as well as
the polarization and temperature dependence of these excitations.
In this picture, the high-energy Raman peak is due to light
scattering from exchange-bond fluctuations, a process which is 
basically identical to magnetic Raman scattering in copper oxides.
A pseudospin-like behaviour of quantum orbitals explains why
the Raman response of two apparently different degrees of freedom 
exhibit very similar features. In fact, the energies of the two-orbiton 
Raman mode in the titanates ($\sim 235$ meV) and the two-magnon mode 
in the cuprates ($\sim 350$ meV) are related by the simple scaling 
relation for the exchange-pair energy $\Omega_{pair}\sim (z-1)J_{SE}$, 
where $J_{SE}$ and $z$ are the strength and number of exchange bonds, 
respectively. These considerations, which can be readily extended to 
other transition metal compounds with more than one valence electron 
(or hole), provide a glimpse of an underlying universality in the 
electronic spectra of these complex materials.

We would like to thank M. Cardona, R. Zeyher, and R. R\"{u}ckamp
for useful discussions.

\end{document}